\documentclass{llncs} 
\usepackage{amssymb}
\usepackage{graphicx}

\usepackage{url}

\usepackage[title]{appendix}
\usepackage{subfig}
\usepackage{algorithm,algpseudocode}

\usepackage{cite}
\usepackage{color}
\usepackage{calc}
\usepackage{siunitx}
\DeclareSIUnit\mt{\milli\tesla} 
\newcolumntype{P}[1]{>{\centering\arraybackslash}p{#1}}

\usepackage[T1]{fontenc}
\usepackage[ansinew]{inputenc}
\usepackage[english]{babel}
\usepackage{adjustbox}
\usepackage{amsmath,amsfonts,amssymb}
\usepackage{xparse}
\usepackage[section]{placeins} 
\usepackage[misc]{ifsym}
\usepackage{url}
\usepackage{caption}

\usepackage[many]{tcolorbox}
\usepackage{tikz}
\usetikzlibrary{decorations.pathreplacing}
\usetikzlibrary{shapes.arrows}
\usetikzlibrary{arrows,shapes,automata,petri,positioning,calc}

\usepackage{array,longtable}

\newcommand{\Poldhn}{\ensuremath{P_\text{HN}} }
\newcommand{\Poldue}{\ensuremath{P_\text{UE}} }
\newcommand{\pis}{p_i}
\newcommand{\di}{d_i}
\newcommand{\pone}{p_1}
\newcommand{\done}{d_1}
\newcommand{\ptwo}{p_2}
\newcommand{\dtwo}{d_2}
\newcommand{\pc}{p_c}
\newcommand{\dc}{d_c}
\newcommand{\pn}{p_n}
\newcommand{\dn}{d_n}
\newcommand{\pf}{p_f}
\newcommand{\df}{d_f}

\newcommand{\I}{\ensuremath{\text{I}}}
\newcommand{\MCC}{\ensuremath{\text{MCC}}}
\newcommand{\MNC}{\ensuremath{\text{MNC}}}
\newcommand{\MSIN}{\ensuremath{\text{MSIN}}}
\newcommand{\SUCI}{\ensuremath{\text{SUCI}}}

\newcommand{\K}{\ensuremath{\text{K}}}
\newcommand{\SQN}{\ensuremath{\text{SQN}}}
\newcommand{\XSQN}{\ensuremath{\text{XSQN}}}

\newcommand{\RAND}{\ensuremath{\text{RAND}}}
\newcommand{\AUTN}{\ensuremath{\text{AUTN}}}
\newcommand{\XRES}{\ensuremath{\text{XRES}}}
\newcommand{\HXRES}{\ensuremath{\text{HXRES}^{*}}}
\newcommand{\RES}{\ensuremath{\text{RES}^{*}}}
\newcommand{\HRES}{\ensuremath{\text{HRES}^{*}}}

\newcommand{\RESLTE}{\ensuremath{\text{RES}}}
\newcommand{\KASME}{\ensuremath{\text{K}_{\text{ASME}}}}
\newcommand{\KSEAF}{\ensuremath{\text{K}_{\text{SEAF}}}}
\newcommand{\MAC}{\text{MAC}}

\newcommand{\Null}{\ensuremath{\mathit{NULL}}}
\newcommand{\salt}{\ensuremath{\mathit{salt}}}

\newcommand{\pk}{\mathit{pk}}
\newcommand{\sk}{\mathit{sk}}

\newcommand{\HNID}{\ensuremath{\text{HNID}}}
\newcommand{\HNPKI}{\ensuremath{\text{HNPKI}}}
\newcommand{\SPSI}{\ensuremath{\text{SUPIPSI}}}

\newcommand{\COUNTER}{\ensuremath{\mathit{CTR}}}
\newcommand{\ECF}{\ensuremath{\mathit{ECF}}} 

\newcommand{\deltamin}{\ensuremath{\delta_{\text{min}}}} 
\newcommand{\deltamax}{\ensuremath{\delta_{\text{max}}}}

\newcommand{\AKAFOURG}{\ensuremath{\texttt{LTEAKA}}}
\newcommand{\AKAFIVEG}{\ensuremath{\texttt{5GAKA}}}
\newcommand{\AVX}{\ensuremath{\texttt{AV}}}
\newcommand{\AVR}{\ensuremath{\texttt{AVR}}}
\newcommand{\LUX}{\ensuremath{\texttt{LU}}}
\newcommand{\AKASUCCEDEDP}{\ensuremath{\texttt{5GAKAS}}}

\newcommand{\UPUE}{\ensuremath{\texttt{UPUE}}}
\newcommand{\UPHN}{\ensuremath{\texttt{UPHN}}}

\definecolor{mygray}{gray}{0.4}

\newcommand{\Encpk}{\ensuremath{\mathcal{E}}}

\expandafter\def\expandafter\UrlBreaks\expandafter{\UrlBreaks
  \do\a\do\b\do\c\do\d\do\e\do\f\do\g\do\h\do\i\do\j%
  \do\k\do\l\do\m\do\n\do\o\do\p\do\q\do\r\do\s\do\t%
  \do\u\do\v\do\w\do\x\do\y\do\z\do\A\do\B\do\C\do\D%
  \do\E\do\F\do\G\do\H\do\I\do\J\do\K\do\L\do\M\do\N%
  \do\O\do\P\do\Q\do\R\do\S\do\T\do\U\do\V\do\W\do\X%
  \do\Y\do\Z\do?\do.\do&}

\begin{document}

\mainmatter  

\title{Defeating the Downgrade Attack on Identity Privacy in 5G}


%
%
\author{Mohsin Khan$^\text{1,2(\Letter)}$%
\and Philip Ginzboorg$^\text{3,4}$
\and Kimmo J\"arvinen$^\text{1,2}$
\and Valtteri Niemi$^\text{1,2}$
}  %



\institute{$^\text{1}$University of Helsinki, Helsinki, Finland\\
$^\text{2}$Helsinki Institute for Information Technology, Helsinki, Finland\\
\{\email{mohsin.khan, kimmo.u.jarvinen, valtteri.niemi}\}\email{@helsinki.fi}\\
$^\text{3}$ Huawei Technologies, Helsinki, Finland\\
$^\text{4}$ Aalto University, Espoo, Finland\\
\email{philip.ginzboorg@huawei.com}
}

%
%

\maketitle

\begin{abstract}
3GPP Release 15, the first 5G standard, includes protection of user identity privacy against IMSI catchers. These protection mechanisms are based on public key encryption. Despite this protection, IMSI catching is still possible in LTE networks which opens the possibility of a downgrade attack on user identity privacy, where a fake LTE base station obtains the identity of a 5G user equipment. We propose (i) to use an existing pseudonym-based solution to protect user identity privacy of 5G user equipment against IMSI catchers in LTE and (ii) to include a mechanism for updating LTE pseudonyms in the public key encryption based 5G identity privacy procedure. The latter helps to recover from a loss of synchronization of LTE pseudonyms. Using this mechanism, pseudonyms in the user equipment and home network are automatically synchronized when the user equipment connects to 5G. Our mechanisms utilize existing LTE and 3GPP Release 15 messages and require modifications only in the user equipment and home network in order to provide identity privacy. Additionally, lawful interception requires minor patching in the serving network. 
\end{abstract}


\section{Introduction} \label{sec:intro}
A generic mobile network has three main parts: (i) user equipment (UE); (ii) home network (HN), i.e. the mobile network where the user has a subscription; and (iii) serving network (SN), i.e. a mobile network that the UE connects to in order to avail services.\footnote{The notation used in this paper is summarized in Appendix \ref{appendix:notations}.} The UE includes mobile equipment (ME) and a universal integrated circuit card (UICC). 
The UE is said to be roaming when the SN and HN are different. 
When both the UE and HN are made to 5G specifications, it is possible that the 5G UE connects to a legacy serving network, e.g., the LTE (long term evolution, or 4G) serving network. It is important to allow such a ``downgraded'' connection, because -- especially in the early phases of 5G adoption -- 5G coverage will be spotty compared to LTE.

International mobile subscriber identity (IMSI) is a globally unique identity of a mobile user. Identity privacy means that long-term user identities remain unknown to everyone else besides UE, SN, and HN. IMSI catchers, i.e. malicious devices that steal IMSIs in order to track and monitor the mobile users, are a threat to users' identity privacy \cite{washingtonpost2014,pki_umts_imsi_catcher,catch_me_if_you_can,pets2017}. 
Passive IMSI catchers attack by eavesdropping; active IMSI catchers attack by impersonating an SN. Mobile networks had protection against passive IMSI catchers since GSM, but until 5G they did not protect users against active IMSI catchers. 

Third generation partnership project (3GPP) Release 15, the first 5G standard, includes protection against active IMSI catchers~\cite{TS22261}. This protection is implemented so that the UE encrypts its identity using public key encryption with the public key of the HN~\cite{TS33501}. The concealed identity is called subscription concealed identifier $(\SUCI)$. This protection works only when the SN is also a 5G entity because an SN from LTE, 3G or GSM networks would not know how to process the $\SUCI$. This implies that an active IMSI catcher can mount a downgrade attack against a 5G UE so that it impersonates an LTE SN and exploits LTE's weakness in order to steal the IMSI of the 5G UE. In this paper, we propose mechanisms to prevent this downgrade attack. 

Our solution uses pseudonyms that have the same format as IMSI for LTE communication to defeat the downgrade attack. The idea of using pseudonyms in IMSI format  to confound 
IMSI catchers in mobile networks has been studied in several works during the recent years \cite{CCS15,SSR15,Norrman_Naslund_Dubrova_2016,Ginzboorg_Niemi_2016,wisec17,ICISS2017}. We take a similar approach as proposed in these papers. A pseudonym looks like a normal IMSI, but its last nine to ten digits are randomized and frequently changing.\footnote{The first five to six digits of the IMSI identify the country and the home network of the mobile user. Even though these digits allow linkability in certain cases, (e.g., if in a visited network there is only one roaming UE from a specific country), these digits are not randomized, because they are needed to route initial requests for authentication data for roaming UE to the correct home network.} The UE is provisioned with two pseudonyms in the beginning and it gets fresh pseudonyms during AKA runs. It uses these pseudonyms instead of IMSI to identify itself when connecting to an LTE SN. The UE uses the SUCI instead of IMSI to connect to a 5G SN. Our solution piggybacks on existing messages involved in the LTE and 5G authentication and key agreement (AKA) protocols to deliver new pseudonyms to the UE and does not require additional messages.

Since the space of pseudonym having IMSI format is limited, pseudonyms need to be reused, i.e., disassociated from one user and reallocated to another. On the other hand, a 5G UE may connect to multiple SNs simultaneously \cite{TS33501}; causing the UE to use different pseudonyms to connect to different SNs. Hence, it is a challenge for the HN to know when it can disassociate a pseudonym from a current user and reallocate it to a new user. In our solution the UE embeds information about its LTE pseudonyms into the SUCI. This enables efficient reuse of pseudonyms in the HN and accurate billing. The UE does not need a pseudonym to connect to a 5G SN. Hence,  even in the unlikely event, where the UE and HN lose synchronization of pseudonyms, the synchronization can be restored because the UE obtains a new LTE pseudonym from the HN simply by connecting to a 5G SN. 

This paper advances protocol design, beyond what is described in papers \cite{CCS15,SSR15,Norrman_Naslund_Dubrova_2016,Ginzboorg_Niemi_2016,wisec17,ICISS2017},  as follows:
\begin{itemize}
 \item[(i)] The HN can reuse pseudonyms even when a UE has simultaneous connections to multiple SNs.
 \item[(ii)] The states of pseudonyms in UE and HN will remain synchronized, given that the HN and the UE function correctly.
 \item[(iii)] For the case where UE and HN get desynchronized due to some unlikely errors, we have detailed a re-synchronization mechanism for pseudonyms' state in UE and HN. The mechanism is based on running a 5G AKA with a 5G SN.
\end{itemize}

Any enhancement of user identity privacy must support Lawful Interception (LI), i.e. selective interception of individual subscribers by legally authorized agencies. In Section \ref{sec:lawful_intercpetion_patching} we describe how LI is supported in the Release 15 enhancement to user identity privacy (where UE encrypts its identity) and propose to supplement pseudonym-based solution with similar features. Adding support for LI into our pseudonym-based solution requires software update (patching) in the core network elements of the LTE SN; it does not require patching in the radio access network elements (base stations).

In this paper we consider the LTE SN in detail. Same solution can be adapted for 3G SNs in a straightforward manner. Its adaptation to GSM would require additional measures because GSM lacks means to authenticate any message received by a mobile device. An example of such a measure has been described in \cite{CCS15}.

\section{Preliminaries} \label{sec:preli}
An IMSI is a globally unique number, usually of 15 decimal digits, identifying a subscriber in GSM, UMTS and LTE systems. The first 3 digits of an IMSI represent the mobile country code (MCC); the second 2 or 3 digits represent the mobile network code (MNC); and the last 10 or 9 digits represent the mobile subscription identification number (MSIN) \cite{TS23003}. The long-term subscriber identifier in 5G system is called Subscription Permanent Identifier (SUPI) \cite{TS23501}. It may contain an IMSI, or a network-specific identifier for private networks. 



\begin{figure}[!t]
\begin{centering}
    \input{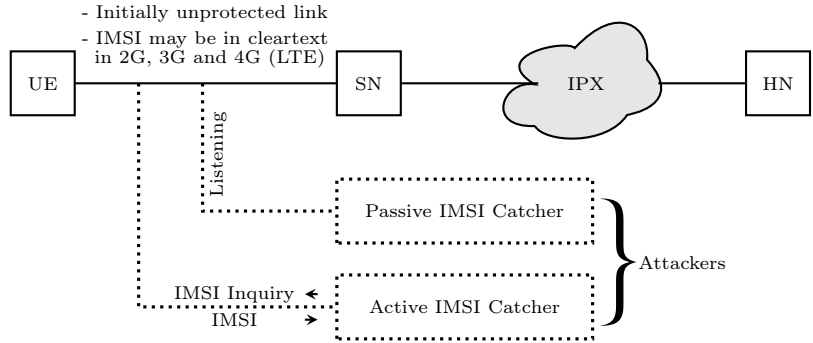}
    \caption{Schematic illustration of mobile network. }
    \label{fig:mobile_network}
\end{centering}    
\end{figure}

Figure \ref{fig:mobile_network} illustrates a high-level architecture of mobile networks. The UE of a user, which has  a subscription with the HN (home network), connects with the SN (serving network) to get services. If the SN and the HN are different networks, we say that the UE is roaming. In that case, the SN and the HN connect with each other over the IP Exchange (IPX) network. The link between the UE and the SN is initially unprotected in LTE, 3G and GSM networks. In 5G, some information sent over this initial link may be confidentiality protected using the public key $\pk$ of the HN. Information needed for routing in IPX cannot be confidentiality protected (otherwise, routing would not work) \cite{fruct2018}.

The HN stores the $(\text{IMSI}, \K)$ pairs of all its subscribers in the subscription database. The 5G HN also has a public/private key pair $\pk,\sk$. The UE includes an ME and a UICC. The UICC is tamper resistant: a malicious entity can not read its content without sophisticated instruments. The universal subscriber identity module (USIM) is an application that runs in the UICC. 

In LTE, the USIM stores the IMSI and the subscriber-specific master key $\K$ of the user. Please note that if a UE conceals the IMSI using $\K$, the HN would not know which $\K$ to use to decrypt the message. Moreover, the UE is not provisioned with any SN-specific keys. Thus, the IMSI is initially sent unprotected over the link between the UE and the SN in LTE. Also in 3G and 2G the IMSI is initially sent unprotected over the link between the UE and the SN. But in 5G the USIM may also store the public key $\pk$ of the HN; and then the UE, before identifying itself, can send encrypted (by the key $\pk$) message to the HN.

When a UE wants to connect to a SN, the SN wants to know the identity of the user so that the user can be billed. The SN sends an IMSI inquiry to the UE. Since the link between the UE and the LTE (also 3G and 2G) SN is initially unprotected, the UE has to respond with its IMSI in cleartext. A passive IMSI catcher listens to the radio channel and waits for IMSIs sent in cleartext. An active IMSI catcher impersonates a legitimate SN and makes an IMSI inquiry. The UE has no way to distinguish an active IMSI catcher from a legitimate SN before authenticating the SN. Hence, the UE invariably sends the IMSI to the attacker in cleartext.

The identification is followed by mutual authentication based on challenge and response. We will now outline the basic principle of the authentication protocol. The SN provides the identity of the user to the HN. The HN (i) prepares a random challenge; (ii) computes the expected response to the challenge, an authentication token, and an anchor key; (iii) may compute some other required information. The response, authentication token and anchor key are computed using the master key \K. An authentication token includes information that protects the integrity of the challenge and defeats replay attack. The challenge, response, authentication token, anchor key and also some other relevant information are collectively known as an authentication vector (AV). The HN sends the AV to the SN. 

The SN sends the challenge and the authentication token to the UE. The UE verifies the integrity and freshness of the challenge using the master key \K. If the verification result is positive, the UE computes the expected response to the challenge using the master key \K and sends the response to the SN. If the expected response that the SN received from the HN matches with the response that the UE sent, the authentication is successful. 

The authentication and key agreement protocols of LTE and 5G networks -- known as LTE AKA \cite{TS33401} and 5G AKA \cite{TS33501} -- are based on the above principle. In LTE AKA, if the authentication is successful, and if the UE that participated in the AKA is attaching with the SN for the first time, then the SN sends a location update (LU) message to the HN \cite{ICISS2017,TS23012}. Once a UE is authenticated, the SN gives it a temporary identity -- known as globally unique temporary UE identity (GUTI) -- with confidentiality and integrity protection over a secure channel that has been established based on the authentication. Since then, the UE uses GUTI to identify itself to the SN. The use of GUTI defeats the passive IMSI catchers. However, the use of GUTI does not defeat active IMSI catchers. This is because the UE or the SN may have lost the GUTI, or the UE may visit a new SN.  In either case, the SN would make an IMSI inquiry towards the UE. The UE would have to respond with IMSI in cleartext, otherwise it would be locked out of the network.

Once a UE is attached to an SN, the UE may go to idle mode. When there is downlink data available for an idle UE, the SN would page the UE to wake it up, using the IMSI or the GUTI (of that UE) in the paging message. It is worth mentioning that a new authentication may be in initiated by the SN after UE responds to a paging message. 

\section{Related Work} \label{sec:related_work}

The attack on user identity privacy by active IMSI catchers has been known since GSM, and numerous defence mechanisms against this attack have been proposed in the literature. We identified two major trends in these mechanisms: the first is based on pseudonyms and the second on public key cryptography. Below, we 
list works that we consider to be the most relevant to this paper.


In Herzberg et al.\,\cite{travelling_incognito}, one-time pseudonyms are created by probabilistic encryption of the user's identity using a key that is known only by the HN. The HN sends a set of pseudonyms to the UE when the UE and the HN have a secure communication channel.

Publications proposing cellular network pseudonyms in the same format as IMSI, but with randomized, frequently changing MSIN, include \cite{CCS15, SSR15, Ginzboorg_Niemi_2016, Norrman_Naslund_Dubrova_2016}. In Broek et al. \cite{CCS15} and Khan and Mitchell \cite{SSR15} the pseudonym's update is embedded in a random challenge, RAND, of AKA. Khan and Mitchell \cite{wisec17} identified a weakness in solutions proposing cellular network pseudonyms in the same format as IMSI, which could be exploited to desynchronize pseudonyms in the UE and the HN. They also proposed a fix. Khan et al. \cite{ICISS2017} found a weakness in \cite{wisec17} and proposed a solution. We will refer to the solution in \cite{ICISS2017} as the KJGN scheme.


Asokan \cite{Asokan94} described how public key encryption can be used to achieve identity privacy in mobile environments. In this solution, only the HN has a public/private key pair and the UE is provisioned with the public key of the HN. The UE encrypts identity information using public key before sending it to the HN.  K{\o}ien \cite{koienibe} suggests using identity based encryption (IBE) to defeat IMSI catchers in LTE. Khan and Niemi \cite{NSS17} propose the use of IBE to defeat IMSI catchers in 5G networks.



5G is the first generation of mobile networks that includes protection against active IMSI-catchers. 3GPP has decided that users' identities will be protected in 5G by including public key encryption~\cite{TS33501}. 
The idea is similar to Asokan~\cite{Asokan94}. In addition, the possibility to include IMSI in the paging message has been removed.
We will now outline the working of 5G protection mechanism \cite{TS33501}.

The UE conceals the 5G user's long-term identities with Elliptic Curve Integrated Encryption Scheme (ECIES)~\cite{secg_sec1,secg_sec2} before sending them to the SN. ECIES is a hybrid encryption scheme that combines an elliptic curve based public key cryptography with secret key cryptography; it is a semantically secure probabilistic encryption scheme ensuring that successive encryptions of the same plaintext with the same public key result in different ciphertexts with very high probability.  Specifically, \cite{TS33501} includes two ECIES profiles, both for the approximately 128-bit security level. Both profiles use AES-128~\cite{fips197} in CTR mode~\cite{sp80038a} for confidentiality and HMAC-SHA-256~\cite{hmac,fips1804} for authenticity in the secret key cryptography part but use either Curve25519~\cite{curve25519,rfc7748} or secp256r1~\cite{secg_sec2} elliptic curves for the public key cryptography part.

A UE that is provisioned with the HN's public key $\pk$,\footnote{The standard \cite{TS33501} does not require the HN to provision $\pk$ into every UE. If HN has not provisioned its $\pk$ into a UE, then that UE will not conceal its long-term identity with this mechanism.} uses it to construct \SUCI\ by computing $\Encpk_{\pk}(\MSIN)$---the encryption of \MSIN\ with the HN's public key $\pk$---and concatenating it with certain cleartext information. As defined in~\cite{TS33501}, this cleartext information includes \HNID, the home network identifier enabling successful routing to the HN, \SPSI\ for defining the scheme (i.e., either a null scheme or the ECIES profile), and \HNPKI\ for denoting which public key was used in encryption. ECIES guarantees that \MSIN\ can be decrypted from \SUCI\ only by the HN who holds the private key corresponding to $\pk$. Hence, the users' identities are protected.

Please note that public-key based solutions, where the IMSI is delivered to the network encrypted by the public key of the network, are conceptually simpler than pseudonym-based solutions, because the latter need a mechanism for changing pseudonyms while keeping the set of pseudonyms in the UE and the HN synchronized. On the other hand, the impact of pseudonyms in IMSI format on legacy network nodes between UE and HN is smaller than of IMSI encrypted by a public key of the network, because the format of IMSI encrypted by a public key of the network is quite different from plaintext IMSI. 



\section{Our Solution} \label{sec:our_solution}
The 5G standard protects long-term user identity against active IMSI catchers by using SUCI (generated by the public key of the HN) \cite{TS33501}, as outlined in Section \ref{sec:related_work}. Despite this protection in 5G, a fake LTE SN can still mount a downgrade attack on the identity privacy of a 5G UE. 

\begin{figure}[!t]
    \centering
    \begin{tikzpicture}[auto, every text node part/.style={align=left}]
[
    pre/.style={=stealth',semithick},
    post/.style={->,shorten >=1pt,>=stealth',semithick}
]

\node (S0) [rectangle, draw = white] {};
\node (dummy1) [rectangle, draw = white,right of = S0,node distance = 9.5cm] {};

\coordinate (S0) at (8,0);
\coordinate (dummy) at (9.5,0);


\node (S1) [rectangle, draw, fill=gray, fill opacity = .05, text opacity=1, right of = S0, node distance = 1.5cm] {
$(\pone,\done)$
};

\node (S2) [rectangle, draw, fill=gray, fill opacity = .01, text opacity=1, right of = S1, node distance = 1.5cm] {
$(\ptwo,\dtwo)$
};

\node (S3) [rectangle, draw, fill=gray, fill opacity = .01, text opacity=1, left of = S1, node distance = 5.5cm] {
$\Poldue = \{(\pis,\di) \vert \text{ where } \di < \done\}$
};

\draw node[left of = S3, node distance = 3.5cm] {\textbf{UE Side:} };

\draw node[above of = S1, node distance = .5cm] {$\done < \dtwo $};


\node (S4) [rectangle, draw, fill=gray, fill opacity = .01, text opacity=1, below of = S1, node distance = 2cm] {$(\pn,\dn)$};

\node (S5) [rectangle, draw, fill=gray, fill opacity = .01, text opacity=1, right of = S4, node distance = 1.5cm] {$(\pf,\df)$};

\node (S5) [rectangle, draw, fill=gray, fill opacity = .01, text opacity=1, left of = S4, node distance = 1.5cm] {$(\pc,\dc)$};

\node (S6) [rectangle, draw, fill=gray, fill opacity = .01, text opacity=1, left of = S5, node distance = 4.0cm] {
$\Poldhn = \{(\pis,\di) \vert \text{ where } \di < \dc\}$
};

\draw node[left of = S6, node distance = 3.5cm] {\textbf{HN Side:} };

\draw node[above of = S4, node distance = .5cm] {$\dc < \dn < \df $};

\end{tikzpicture}
    \caption{Pseudonyms in UE and HN} \label{fig:pseudonyms_in_UE_HN}
\end{figure}
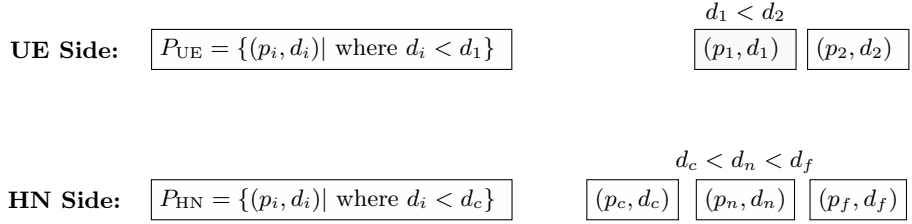

We mitigate the downgrade attack as follows: instead of IMSI, a 5G UE  in LTE network uses pseudonyms that have the same format as IMSI. In addition, when a 5G UE runs 5G AKA, it also synchronizes its LTE pseudonyms with the HN. In our solution:
\begin{enumerate}
    \item a 5G UE uses pseudonyms to connect with LTE SNs and Release-15 SUCI to connect with 5G SNs;
    \item only the HN allocates and releases pseudonyms of mobile users; initially the HN allocates two pseudonyms per 5G user and provisions them into those users' USIMs;
    \item a 5G UE gets new pseudonyms by participating in authentication protocols: LTE AKA or 5G AKA; the two latest pseudonyms received by the UE during successful AKA are denoted by $\pone$ and $\ptwo$;
    \item in order to support simultaneous connections with several SNs, our solution: 
        \begin{itemize}
            \item[(i)] uses subscriber-specific counter $d$ of pseudonyms maintained by the HN;
            \item[(ii)] keeps track of in-use pseudonyms in 5G UE and HN, using sets $\Poldue$ and $\Poldhn$, respectively; the elements of these sets are pairs $(\pis,\di)$ of pseudonyms and their respective counters (see Figure \ref{fig:pseudonyms_in_UE_HN}); 
            \item[(iii)] piggybacks information about pseudonyms in 5G UE within the Release-15 SUCI.
        \end{itemize}
\end{enumerate}


The 5G UE will use only $\pone$ or $\ptwo$ when replying to IMSI inquiry from an LTE SN. The set $\Poldue$ contains pseudonyms that 5G UE received before $\pone$ and $\ptwo$. The UE deletes pseudonyms from $\Poldue$ based on policy provided by the HN that could include, e.g., pseudonyms' lifetime or the maximum size of $\Poldue$. The $\Poldhn$ contains pseudonyms that the HN thinks are in $\Poldue$, and it deletes pseudonyms from $\Poldhn$ according to another policy. One objective of these policies is that the HN should not delete a pseudonym which the UE has not deleted yet. Thus, the pseudonyms in the UE constitute a subset of that UE's pseudonyms in the HN. In short, the UE informs the HN about its oldest pseudonym when it connects with a 5G SN using SUCI, and then the HN is able to reduce the set $\Poldhn$. Please note that as long as the UE is connecting to LTE SN only, the size of $\Poldhn$ grows. It will be explained later how to avoid that $\Poldhn$ grows too much.


Our solution does not modify the structure and/or length of any existing message. Also, it does not introduce any new messages on top of what 3GPP has standardized; it only introduces changes in the 5G UE and its HN. However, to enable lawful interception in the SN, we would need some changes in the LTE SNs too. This issue is discussed in Section \ref{sec:lawful_intercpetion_patching}.

As standardized, a 5G USIM comes with an IMSI, a master key $\K$ and the HN's public key $\pk$ embedded in it. A 5G USIM in our solution also has to include two pseudonyms $\pone,\ptwo$ and a key $\kappa$, shared with HN, for decrypting the pseudonyms. 
Similarly, along with the user's IMSI, and master key $\K$, the HN in our solution has to store additional information: the shared key $\kappa$ for encrypting pseudonyms and three pseudonyms $\pc,\pn, \text{ and } \pf$ (where the subscripts stand for ``current,'' ``next'' and ``future''). Ideally $\pone = \pc$ and $\ptwo = \pn$.


When a pseudonym $p$ is allocated to a subscriber, it is associated with a subscriber-specific counter $d$. We require that $d$ is a strictly monotonically increasing integer variable that increases each time the HN allocates a new pseudonym to the subscriber. 





\subsection{LTE AKA Based Solution} \label{sec:lte_aka_based_solution}
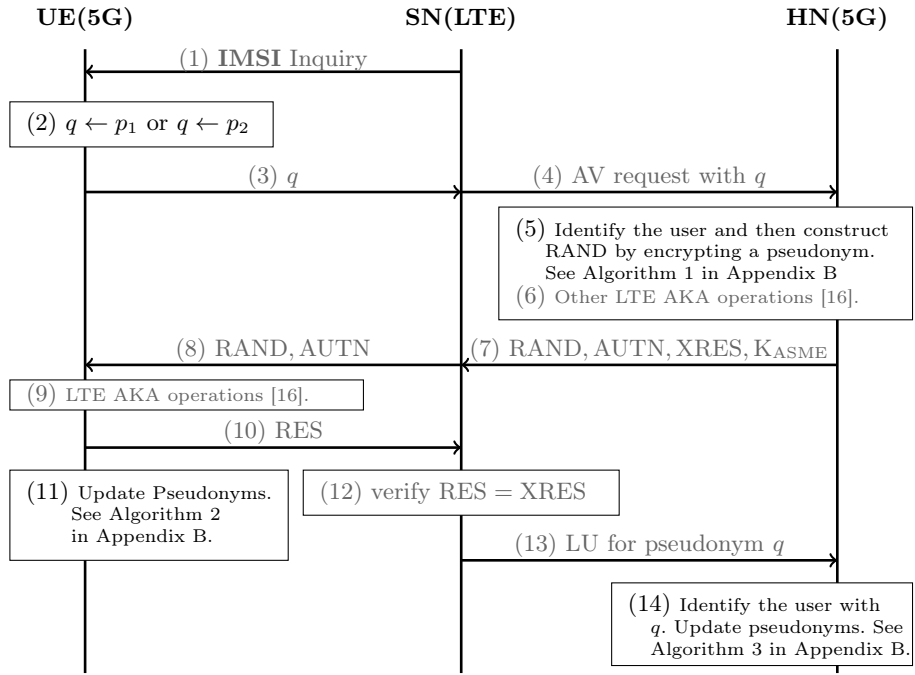
\begin{figure}[!t]
    \centering
    \begin{tikzpicture}[scale=1]


\draw node at (11,11.4) {\textbf{HN(5G)}};
\draw [black, line width = 1pt] (11,11) -- (11,2.7);

\draw node at (6,11.4) {\textbf{SN(LTE)}};
\draw [black, line width = 1pt] (6,11) -- (6,2.7);

\draw node at (1,11.4) {\textbf{UE(5G)}};
\draw [black, line width = 1pt] (1,11) -- (1,2.7);

\draw [black, line width = 1pt, ->] (6.0,10.7) -- (1,10.7);
\draw node at (3.5,10.85) { \textcolor{mygray}{(1) \textbf{IMSI} Inquiry}};

\filldraw[draw=black,fill=white] (0,9.7) rectangle (3.5,10.3);
\draw node[anchor = west] at (0.1,10) { (2) $q \leftarrow p_{1}$ or $q \leftarrow p_{2}$};

\draw [black, line width = 1pt, <-] (6,9.1) -- (1,9.1);
\draw node at (3.5,9.3) {\textcolor{mygray}{(3) $q$}};

\draw [black, line width = 1pt, ->] (6,9.1) -- (11,9.1);
\draw node at (8.5,9.3) {\textcolor{mygray}{(4) AV request with $q$}};

\filldraw[draw=black,fill=white] (6.5,7.4) rectangle (12,8.9);
\draw node[anchor = west] at (6.6,8.6) {(5) \scriptsize Identify the user and then construct};
\draw node[anchor = west] at (7,8.3) {\scriptsize $\RAND$ by encrypting a pseudonym.};
\draw node[anchor = west] at (7,8) {\scriptsize See Algorithm \ref{alg:conctruct_rand_lte} in Appendix \ref{appendix:algorithms}};

\draw node[anchor = west] at (6.6,7.7) {\textcolor{mygray}{(6) \scriptsize Other LTE AKA operations \cite{TS33401}.}};

\draw [black, line width = 1pt, <-] (6,6.8) -- (11,6.8);
\draw node at (8.5,7) {\textcolor{mygray}{(7) $\RAND,\AUTN,\XRES,\KASME$}};

\draw [black, line width = 1pt, ->] (6,6.8) -- (1,6.8);
\draw node at (3.5,7) {\textcolor{mygray}{(8) $\RAND,\AUTN$}};

\filldraw[draw=black,fill=white] (0,6.2) rectangle (4.7,6.6);
\draw node[anchor = west] at (0.1,6.4) {\textcolor{mygray}{(9) \scriptsize LTE AKA operations \cite{TS33401}. }};

\draw [black, line width = 1pt, <-] (6,5.7) -- (1,5.7);
\draw node at (3.5,5.9) {\textcolor{mygray}{(10) $\RESLTE$}};

\filldraw[draw=black,fill=white] (3.9,4.8) rectangle (8.1,5.4);
\draw node[anchor = west] at (4.0,5.1) {\textcolor{mygray}{(12) verify $\RESLTE = \XRES$}};

\draw [black, line width = 1pt, ->] (6,4.2) -- (11,4.2);
\draw node at (8.5,4.4) {\textcolor{mygray}{(13) LU for pseudonym $q$}};

\filldraw[draw=black,fill=white] (8,2.8) rectangle (12,3.9);
\draw node[anchor = west] at (8.1,3.6) {(14) \scriptsize Identify the user with };
\draw node[anchor = west] at (8.4,3.3) {\scriptsize $q$. Update pseudonyms. See};

\draw node[anchor = west] at (8.4,3.0) {\scriptsize  Algorithm \ref{alg:update_pseudonym_hn} in Appendix \ref{appendix:algorithms}.};

\filldraw[draw=black,fill=white] (0,4.2) rectangle (3.7,5.4);
\draw node[anchor = west] at (0.1,5.1) {(11) \scriptsize Update Pseudonyms.};
\draw node[anchor = west] at (0.7,4.8) {\scriptsize See Algorithm \ref{alg:extract_pseudonym} };
\draw node[anchor = west] at (0.7,4.5) {\scriptsize in Appendix \ref{appendix:algorithms}.};

\end{tikzpicture}
    \caption{Solution, when the SN is from an LTE network.}
    \label{fig:lte_based_solution}
\end{figure}

This solution is shown in Figure \ref{fig:lte_based_solution}. It is built on top of the KJGN scheme, which was built on top of LTE AKA. The differences to LTE AKA are indicated by darker font in the figure. The additions on top of KJGN scheme include the use of the counter $d_i$, which is associated with pseudonym $p_i$, and the sets $\Poldue,\Poldhn$. One major modification is in the condition on which a UE or the HN forget pseudonyms.  Supplements for lawful interception are not shown in Figure \ref{fig:lte_based_solution}; they will be discussed separately in Section \ref{sec:lawful_intercpetion_patching}.

\subsubsection{Description.}
\begin{itemize}
    \item[(1)] An LTE SN inquires the UE about the IMSI. 
    \item[(2)] The UE chooses one of the pseudonyms $\pone,\ptwo$ and assigns it to $q$. 
    \item[(3)] The UE sends $q$ to the SN.
    \item[(4)] The SN sends an AV request for the pseudonym $q$ to the HN. 
    It is worth mentioning here that in most of the times the user identifies itself with GUTI. Sometimes the user would implicitly identify itself by responding to a paging message. In either case, if the SN wants to perform an LTE AKA, the SN requests an AV to the HN for the pseudonym/IMSI that was associated with the GUTI or was used in the paging message.
    \item[(5)] The HN checks if the pseudonym $q$ is in use for any subscriber. If yes, the HN starts to prepare the AV. It first constructs the random challenge $\RAND$. A new pseudonym is embedded (encrypted with key $\kappa$) in the $\RAND$. Detail of how $\RAND$ is constructed is presented in Algorithm \ref{alg:conctruct_rand_lte} in Appendix \ref{appendix:algorithms}. We also describe it in the following.  
    
    \begin{itemize} 
        \item The 128-bit long random challenge $\RAND$ is created by encrypting (using key $\kappa$ that can be generated from the master key $\K$) the pseudonym $\pf$, its counter $\df$, an error correction flag ($\ECF$) and a randomly chosen $l$-bit long $\salt$. If the pseudonym $\pf$ is null, a new $m$-bit long $\pf$ is chosen randomly from the pool of unused pseudonyms. Then $\df$ is set to the current value of the counter $\COUNTER$, which is a strictly monotonically increasing counter maintained by the HN. It increases each time the HN generates a new pseudonym. The flag $\ECF$ is by default set to $0$ but a 5G HN may set it to some other values to notify the UE about an error in the UE's pseudonym state.
        \item The value of $l$ is equal to $(128 - \text{length}(\df) - \text{length}(\ECF) - m)$. The value of $m$ depends on how many digits of the IMSI are randomized. Since the number of randomized digits can be at most 10, $m \leq 34$. The length of $\df$ and $ECF$ depends on implementation; $\text{length}(\df) \leq 24$ and $\text{length}(\ECF) \leq 2$ bits should be enough. This implies $l \geq 68$.
    \end{itemize}
     
    \item[(6)] The HN computes other parts of authentication vector AV (in addition to the RAND), e.g., the expected response $\XRES$ to the challenge $\RAND$, anchoring key $\KASME$, and authentication token $\AUTN$ \cite{TS33401,ICISS2017}.
    \item[(7)] The HN sends $\RAND$, $\AUTN$, $\XRES$ and $\KASME$ to the SN. 
    \item[(8)] The SN forwards $\RAND$ and $\AUTN$ to the UE.
    \item[(9)] The UE performs LTE AKA related operations, e.g., verifying $\MAC$ in $\AUTN$, computing response $\RES$ \cite{TS33401,ICISS2017}.
    \item[(10)] If everything is fine in Step (9), the UE sends $\RESLTE$ to the SN.
    \item[(11)] The UE decrypts $\RAND$ to extract embedded pseudonym $p$, and the counter $d$; and updates the pseudonyms $\pone,\ptwo$ if $p$ is new. These operations are presented in Algorithm \ref{alg:extract_pseudonym} in Appendix \ref{appendix:algorithms}, and also described in the following.
    \begin{itemize}
        \item UE decrypts $\RAND$ using key $\kappa$ and gets $p,d,\ECF$ and $\salt$ (Line \ref{state:decrypting_rand}).
        \item In an LTE AKA, the $\ECF$ bit is always set to 0.
        \item If the pseudonym $p$ is a new pseudonym i.e., $d > \dtwo$, then the UE inserts $(\pone,\done)$ into $\Poldue$ and sets $(\pone,\done),(\ptwo,\dtwo) \gets (\ptwo,\dtwo),(p,d)$. See lines through \ref{state:p_not_in_p1_p2}-\ref{state:end_inner_if}. 
        \item If $d \leq \dtwo$, then $p$ is considered as an old pseudonym and the UE does not update pseudonyms.
            \begin{itemize}
                \item[--] If somehow the value of $\dtwo$ (in the UE) gets corrupted and becomes larger than $\df$ (in the HN), the UE would never be able to accept new pseudonyms anymore just by running LTE AKA. But even then the UE can still obtain a new pseudonym by running a 5G AKA.
            \end{itemize}
        
    \end{itemize}

    \item[(12)] The SN compares $\RESLTE$ and $\XRES$; the SN stops if $\RESLTE \neq \XRES$.
    \item[(13)] The SN sends an LU message to the HN for the pseudonym $q$.
    \item[(14)] The HN searches for a user s.t., $q \in \{\pn,\pf\}$. If found, and $\pf$ is not null, then the HN inserts $(\pc,\dc)$ into $\Poldhn$ and sets $(\pc,\dc),(\pn,\dn), (\pf,\df) \gets (\pn,\dn), (\pf,\df), (\Null,\Null)$. See Algorithm \ref{alg:update_pseudonym_hn} in Appendix \ref{appendix:algorithms}.
\end{itemize}


\subsection{5G AKA Based Solution} \label{sec:5G_AKA_based_solution}
This solution is used for: (i) delivering a new pseudonym to a 5G UE using 5G AKA; (ii) notifying the HN about pseudonyms that the UE is not using anymore; so that those pseudonyms can be reused in HN; and (iii) re-synchronization of pseudonym states in the (rather unlikely) erroneous situation where $\dtwo$ becomes greater than $\df$. It is required to deliver new pseudonyms to a 5G UE even when the UE has not used the existing pseudonyms to connect with a legitimate LTE SN. This is because, the UE may have used those pseudonyms with an active IMSI catcher. If a 5G UE always connects with a 5G SN, and does not get new pseudonyms by participating in 5G AKA, then the 5G UE will have the same pseudonym for long time. So, if an active IMSI catcher makes many IMSI inquiries over this long time, then the UE would respond to each of those IMSI inquiries with the same pseudonym. Thus, an active IMSI catcher would be able to track and monitor the user with this long-lived pseudonym.

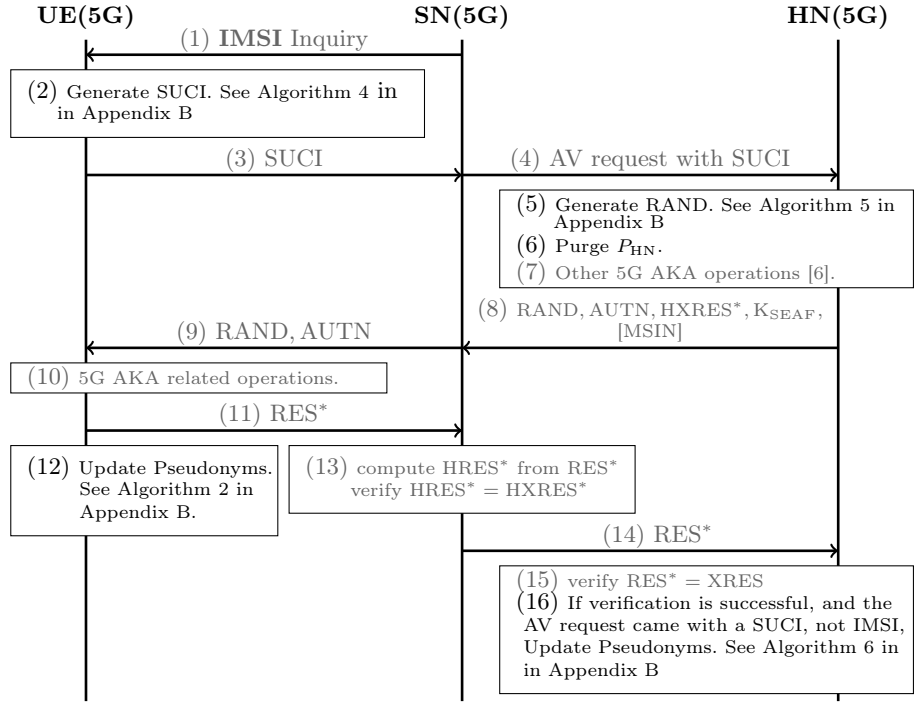
\begin{figure}[!t]
    \centering
    \begin{tikzpicture}[scale=1]



\draw node at (11,11.3) {\textbf{HN(5G)}};
\draw [black, line width = 1pt] (11,11) -- (11,2.2);

\draw node at (6,11.3) {\textbf{SN(5G)}};
\draw [black, line width = 1pt] (6,11) -- (6,2.2);

\draw node at (1,11.3) {\textbf{UE(5G)}};
\draw [black, line width = 1pt] (1,11) -- (1,2.2);

\draw [black, line width = 1pt, ->] (6.0,10.8) -- (1,10.8);
\draw node at (3.5,11.0) { \textcolor{mygray}{ (1) \textbf{IMSI} Inquiry}};

\filldraw[draw=black,fill=white] (0,10.6) rectangle (5.5,9.7);
\draw node[anchor=west] at (0.1,10.3) {\textcolor{black}{(2) \scriptsize Generate SUCI. See Algorithm \ref{alg:generate_suci}} in};

\draw node[anchor=west] at (0.5,10.0) {\scriptsize in Appendix \ref{appendix:algorithms} };

\draw [black, line width = 1pt, ->] (1.0,9.2) -- (6,9.2);
\draw node at (3.5,9.4) {\textcolor{mygray}{(3) $\SUCI$}};

\draw [black, line width = 1pt, ->] (6.0,9.2) -- (11,9.2);
\draw node at (8.5,9.4) {\textcolor{mygray}{(4) AV request with $\SUCI$}};

\filldraw[draw=black,fill=white] (6.5,7.7) rectangle (12,9);
\draw node[anchor=west] at (6.6,8.8) {\textcolor{black}{(5) \scriptsize Generate $\RAND$. See Algorithm \ref{alg:generate_rand_5G} in }};
\draw node[anchor=west] at (7.1,8.55) {\scriptsize Appendix \ref{appendix:algorithms} };
\draw node[anchor=west] at (6.6,8.25) {\textcolor{black}{(6) \scriptsize Purge $\Poldhn$.}};
\draw node[anchor = west] at (6.6,7.9) {\textcolor{mygray}{(7) \scriptsize Other 5G AKA operations \cite{TS33501}.}};

\draw [black, line width = 1pt, <-] (6,6.9) -- (11,6.9);
\draw node at (8.5,7.4) {\textcolor{mygray}{(8) \scriptsize $\RAND,\AUTN,\HXRES,\KSEAF,$}};
\draw node at (8.5,7.1) {\textcolor{mygray}{\scriptsize [$\MSIN$]}};

\draw [black, line width = 1pt, ->] (6,6.9) -- (1,6.9);
\draw node at (3.5,7.1) {\textcolor{mygray}{(9) $\RAND,\AUTN$}};

\filldraw[draw=black,fill=white] (0,6.3) rectangle (5,6.7);
\draw node[anchor = west] at (.1,6.5) {\textcolor{mygray}{(10) \scriptsize 5G AKA related operations.}};

\draw [black, line width = 1pt, <-] (6,5.8) -- (1,5.8);
\draw node at (3.5,6) {\textcolor{mygray}{(11) $\RES$}};

\filldraw[draw=black,fill=white] (0,4.4) rectangle (3.55,5.6);
\draw node[anchor = west] at (0.1,5.3) {\textcolor{black}{(12) \scriptsize Update Pseudonyms.}};
\draw node[anchor = west] at (0.8,5) {\textcolor{black}{\scriptsize See Algorithm \ref{alg:extract_pseudonym} in }};
\draw node[anchor = west] at (0.8,4.7) {\textcolor{black}{\scriptsize Appendix \ref{appendix:algorithms}.}};

\filldraw[draw=black,fill=white] (3.7,4.7) rectangle (8.3,5.6);
\draw node[anchor = west] at (3.8,5.3) {\textcolor{mygray}{(13) \scriptsize compute $\HRES$ from $\RES$}};
\draw node[anchor = west] at (4.4,5) {\textcolor{mygray}{\scriptsize verify $\HRES = \HXRES$}};

\draw [black, line width = 1pt, ->] (6,4.2) -- (11,4.2);
\draw node at (8.6,4.4) {\textcolor{mygray}{(14) $\RES$}};

\filldraw[draw=black,fill=white] (6.5,4) rectangle (12,2.3);
\draw node[anchor = west] at (6.6,3.8) {\textcolor{mygray}{(15) \scriptsize verify $\RES = \XRES$}};
\draw node[anchor = west] at (6.6,3.5) {\textcolor{black}{(16) \scriptsize If verification is successful, and the}};
\draw node[anchor = west] at (6.7,3.2) {\textcolor{black}{ \scriptsize AV request came with a SUCI, not IMSI,}};
\draw node[anchor = west] at (6.7,2.9) {\textcolor{black}{\scriptsize Update Pseudonyms. See Algorithm \ref{alg:update_pseudonym_hn_5G} in }};
\draw node[anchor = west] at (6.7,2.6) {\textcolor{black}{\scriptsize in Appendix \ref{appendix:algorithms} }};

\end{tikzpicture}
    \caption{Solution when the SN is from a 5G network.} \label{fig:5G_AKA_based_solution}
\end{figure}

This solution is built on the 5G AKA protocol of Release 15 \cite{TS33501}; with changes only in the 5G UE and the HN. Thus, the solution is transparent to the 5G SNs of release 15. The solution requires the HN and the USIM to contain all the information the LTE AKA based solution (see Section \ref{sec:lte_aka_based_solution}) requires. Moreover, it requires the HN to have a public/private key pair $\mathit{pk},\mathit{sk}$ and the USIM to be provisioned with the HN's public key $\mathit{pk}$. The solution is presented in Figure \ref{fig:5G_AKA_based_solution}. The changes in 5G AKA are marked by darker texts. 

\subsubsection{Description.}
\begin{itemize}
    \item[(1)] A 5G SN inquires the UE about the IMSI.
    \item[(2)] The UE generates a SUCI. Construction of SUCI by encrypting MSIN using the HN's public key $\mathit{pk}$ is discussed in Section \ref{sec:related_work} and in \cite{TS33501}. We bring changes in the plaintext that is encrypted into SUCI.  Along with MSIN, we also encrypt two counters: $\delta_{\text{min}}$ and $\delta_{\text{max}}$. Here, $\delta_{\text{min}}$ is the smallest counter of all the pseudonyms in the UE. Thus, the HN can know, which pseudonyms the UE is not using anymore; consequently they can be allocated to other UEs. The value of the counter $\delta_{\text{max}}$ is always set to $\dtwo$. Our construction of SUCI is presented in Appendix \ref{appendix:algorithms}, Algorithm \ref{alg:generate_suci}, and also described in the following.
    \begin{itemize}
        \item  UE computes MAC $T$ of message $\mathit{MSIN}||\delta_{\text{min}}||\delta_{\text{max}}$ with master key $\K$ (Line \ref{algo_state:suci_generation_T}). 
        \item UE encrypts $\mathit{MSIN}||\delta_{\text{min}}||\delta_{\text{max}}||T$ with HN's public key $pk$. 
        \item The ciphertext is concatenated with other information that are in plaintext: HN identifier, the public key identifier of the HN, and the SUPI protection scheme identifier (Line \ref{algo_state:suci_generation_encryption}). The outcome is returned as $\SUCI$.
    \end{itemize}
    \item[(3)] The UE sends $\SUCI$.
    \item[(4)] The SN forwards the $\SUCI$ to the HN, requesting an AV. Note that, in most of the times the user identifies itself with GUTI. Sometimes the user would implicitly identify itself by responding to a paging message. In either case, if the SN wants to perform a 5G AKA, the SN requests an AV to the HN with the IMSI that was associated with the GUTI or in the paging.
    \item[(5)] The HN constructs the $\RAND$ by embedding a pseudonym in it. The detail is presented in Algorithm \ref{alg:generate_rand_5G} as described in the following.
    \begin{itemize}
        \item The HN extracts $\MSIN,\deltamin,\deltamax$ and $T$ from the encrypted part of the $\SUCI$ using the private key $\sk$ of the HN  (Line \ref{algo_state:suci_extraction}).
        \item Verifies the MAC $T$ using master key $\K$ (Line \ref{algo_state_T_verification}). If the verification is unsuccessful, the algorithm stops. If the verification is successful, the algorithm continues as following.
        \item Checks if $\pf$ is $\Null$. If yes, an $m$-bit long $\pf$ is randomly allocated (from the pool of free pseudonyms) and $\df$ is set to $\COUNTER$  (Lines \ref{algo_state:allocating_pf_5G}, \ref{algo_state:setting_df_5G}). $\COUNTER$ is a subscriber-specific counter maintained by the HN. It increases every time the HN generates a new pseudonym. 
        \item Sets $\ECF$ to 0. Then checks if $\deltamax$ is greater than $\df$. If yes, it sets $\ECF$ to 1. See Lines from \ref{algo_state:setting_ECF_start} to \ref{algo_state:setting_ECF_end}.
        \item An $l$-bit long random $\salt$ is chosen (Line \ref{algo_state:choosing_salt_5G}).
        \item $(\pf,\df)$ is set into $(p,d)$ i.e., $(p,d)\gets (\pf,\df)$. See line \ref{algo_state:setting_p_d_5G}.
        \item $(p,d,\ECF,\salt)$ is encrypted with key $\kappa$. The ciphertext is $\RAND$. (Line \ref{algo_state:encrypting_into_rand_5G}). It is worth mentioning that the key $\kappa$ can be derived from the master key.
        \item The value of $m$ and $l$ is discussed in Section \ref{sec:lte_aka_based_solution}.
    \end{itemize}
    \item[(6)] HN removes pseudonyms from $\Poldhn$ which have counter smaller than $\deltamin$.
    \item[(7)] HN performs other operations of 5G AKA except constructing $\RAND$ \cite{TS33501}.
    \item[(8)] HN sends an AV ($\RAND,\AUTN,\HXRES,\KSEAF,\MSIN$) to the SN.
    \item[(9)] SN forwards the $\RAND$ and $\AUTN$ to the UE. 
    \item[(10)] UE performs 5G AKA related operations e.g., verifying $\AUTN$ and computing the response $\RES$ to the challenge \cite{TS33501}.
    \item[(11)] UE sends the response $\RES$ to he SN.
    \item[(12)] UE decrypts $\RAND$; extract the embedded pseudonym from the $\RAND$; and updates the pseudonyms in the UE. Algorithm \ref{alg:extract_pseudonym} presents the details. This algorithm is partially described in Section \ref{sec:lte_aka_based_solution}. In 5G AKA the $\ECF$ might be set to $1$ by the HN. In this case the UE would empty the set $\Poldue$, set $(\pone,\done),(\ptwo,\dtwo) \gets (p,d-1),(p,d)$ and terminates the algorithm at this step. This is needed to recover from a very unlikely error situation where $\dtwo$ gets corrupted in the UE. We explain the details in the end of Section \ref{sec:synchronization_alanlysis}.
    \item[(13)] SN computes $\HRES$ as a function of $\RES$; compares $\HRES$ with $\HXRES$.
    \item[(14)] If the comparison in last step matches, SN forwards the $\RES$ to the HN.
    \item[(15)] The HN compares $\RES$ and $\XRES$.
    \item[(16)] If the comparison in previous step matches, HN checks whether the AV (associated with the current current 5G AKA run) came with a SUCI or an IMSI. If with a SUCI, HN checks if the pseudonym $p$ that was embedded in the $\RAND$ is still $\pf$. If yes, the HN moves $(\pc,\dc)$ to $\Poldhn$ and sets $(\pc,\dc),(\pn,\dn), (\pf,\df) \gets (\pn,\dn), (\pf,\df), (\Null,\Null)$. Consequently, the HN would embed a new pseudonym in the $\RAND$ while responding to the next AV request. Details are presented in Algorithm \ref{alg:update_pseudonym_hn_5G} in  Appendix \ref{appendix:algorithms}. On the other hand, if the UE identified itself with GUTI or responded to a paging message, then the subsequent AV request sent by the SN would be with an IMSI, not with a SUCI. Consequently the pseudonyms would not be updated in HN. This means, in response to the next  AV request, the HN will embed the same pseudonym in the $\RAND$. 
\end{itemize}


Step (16) helps the system to avoid generating unnecessary pseudonyms. If a 5G UE attempts to connect with an LTE SN using a pseudonym, but the subsequent LTE AKA fails or no LTE AKA follows (possibly because the SN is an active IMSI catcher), then the next time the UE tries to connect with a 5G SN, it uses SUCI instead of GUTI. In this way the UE can notify the HN that it needs new pseudonym, and it would receive a new pseudonym in the next AKA. Thus, the solution avoids generating unnecessary pseudonyms.

\subsection{Allocation of Pseudonyms} 
In our solution, the pseudonyms are allocated to users by the HN. A pseudonym must not be allocated to two different users at the same time, because such double allocation would hinder correct call routing and billing; e.g,. a user could receive the bill for services used by another user. 
As new pseudonyms are generated for a user, the older pseudonyms are stored in the sets $\Poldhn$ and $\Poldue$. A pseudonym should not be allocated to a new user as long as it is in $\Poldhn$ and $\Poldue$ of any other user. If pseudonyms are never removed from $\Poldhn$ and $\Poldue$, the system will eventually run out of free pseudonyms. Hence, policies are needed for removing pseudonyms from these sets. One objective of these policies is that, a pseudonym should not be deleted from $\Poldhn$ of a user that is not yet deleted from the set $\Poldue$ of that user. 

The HN randomly allocates a new pseudonym for a user from a pool of free pseudonyms, maintained by the HN. A pseudonym $p$ can be in the pool of free pseudonyms only if it is not in the set $\Poldhn$ (or used as $\pc$, $\pn$, or $\pf$) for any user. To keep the pool of free pseudonyms large enough, the HN needs to remove old pseudonyms that are not anymore used by a UE from $\Poldhn$. Before removing a pseudonym $p$ from $\Poldhn$, the HN needs to know that the UE is no longer using it and has removed it from the corresponding $\Poldue$. 

In our solution, a UE removes pseudonyms from $\Poldue$ according to the policies provisioned in the UE by the HN. The UE notifies the HN about the pseudonyms that the HN can remove from the $\Poldhn$. The UE sends (encrypted in the SUCI message) the smallest counter value $\deltamin$ of all the pseudonyms available in the UE. The HN then removes (from $\Poldhn$) all pseudonyms that have smaller counter values than $\deltamin$. The UE sends $\deltamin$ with both integrity and confidentiality protection within the encrypted part of $\SUCI$, as discussed in Section~\ref{sec:5G_AKA_based_solution}.

As mentioned earlier, as long as the UE is connecting to LTE SN only, the size of $\Poldhn$ grows. We will explain next how to avoid that $\Poldhn$ grows too much. The HN needs to have a cap for the size of $\Poldhn$. If the cap is reached, the HN would not generate any future pseudonym $\pf$ for the user --- but will always embed the same pseudonym $\pn$ in the $\RAND$. As a consequence, the user does not enjoy full identity privacy before it connects to a 5G SN using SUCI; the SUCI would include $\deltamin$ and the HN would be able to reduce $\Poldhn$. The details of using this cap are left for further study. The cap should be large enough so that it takes a reasonably long time for a UE to reach the cap; on the other hand the cap should be small enough so that a set of legitimate but malicious UE can not exhaust the pseudonym space by connecting to LTE SNs many times. Another topic for future work is how to maximize the chance that the UE would connect with a 5G SN using SUCI. For example, the UE may connect to 5G in order to reduce the size of $\Poldue$, even if the user is not paying for 5G services. Another way to recover is to send a message to the user about the situation; suggesting to connect to a 5G SN to enjoy better privacy. Other, more sophisticated techniques than capping the size of $\Poldhn$ can also be in the scope for further study.

It is important to define when the UE can decide that it no longer uses a pseudonym and it can be removed from $\Poldue$. The pseudonyms in $\Poldue$ are stored because the UE should be able to respond to paging messages sent by the SN. Therefore, if a UE has a pseudonym (and the associated GUTI) that has not been used for a reasonably  long time (as defined in the policy) and the UE is currently connected to a different SN, the pseudonym can be removed.

The UE may have an old pseudonym in $\Poldue$ that is associated with a GUTI and a security context but has no other pseudonyms associated with the same SN and the UE is currently connected to this SN. In such a case, the UE would initiate a new registration procedure with the SN using pseudonym  $\pone$ or $\ptwo$. If this registration is successful, the UE can remove the old pseudonym from $\Poldue$. The UE may also follow a guideline set by the HN to remove pseudonyms from $\Poldue$; e.g., remove pseudonyms that are older than one day. 

\subsubsection{How Large the Sets $\Poldue$ and $\Poldhn$ Can be.}
In the HN, if 50 \% of MSIN space is used by normal IMSI and pseudonyms, and the allocation of a new pseudonym in HN involves (i) generating a random MSIN, and (ii) checking if that MSIN value has been already allocated, then on the average it would take two tries (each try consisting of (i) and (ii)) for the HN to find a free MSIN. Because of efficiency reasons, we do not want to exceed this number of tries on the average. It follows that the target number of tries and the current level of MSIN space allocation determine the maximum for the average number of pseudonyms per user. Thus, the policy of handling the size of $\Poldue$ has to be adjusted so that the average size of $\Poldue$ would not exceed a certain limit, which in the worst case could be in the order of ten. 

We estimate that only a small fraction of MSIN's space, in the order of 1 \% or less, is in currently in use in the major mobile networks. The biggest fraction, roughly 3 \%, of the MSIN's space seems to be in use in China Mobile networks, which has three MNC codes \cite{mccmnc} and 910 million subscribers \cite{chinamobilesubscribers}. If the average size of $\Poldhn$ is ten, then for each subscriber, around $13$ (also $\pone,\ptwo$ and IMSI) elements from the MSIN space would be allocated on the average. Therefore, the fraction of in use MSIN's space will grow by a factor of 13, e.g., from 3 \% to $39$ \%.

\subsubsection{Alternative Allocation Mechanisms.}
In our solution, pseudonyms are allocated to users in the HN. Another approach would be to generate pseudonyms on the UE side. However, in this other approach, the HN has to be able to map a pseudonym with the correct IMSI of the user. Here we briefly present few alternative options and their downsides.

In this option, the UE may perform format preserving encryption (FPE) of the MSIN with a shared symmetric key and use the ciphertext along with MCC and MNC to construct the pseudonym. There may be only one shared key for the whole network, or separate shared keys for each small group of users. Only one key for the whole network is not secure because an attacker can easily know the key. Indeed, the attacker would only need to be be a valid subscriber of the HN and the ability to read the UICC. On the other hand, separate shared keys for each small group do not result into good privacy. If the size of the group is $k$, the subscriber can achieve only $k$-anonymity. This is because the user would need to identify the user's group in plaintext. In the roaming case, it could be even worse than $k$-anonymity. For instance, it could be the case that only one member of the group is roaming in a certain country at a certain time point.

Another option is that the UE pseudorandomly generates pseudonyms --- e.g., by hashing the IMSI and a salt using the shared master key. The UE uses $m$ bits of the hash digest along with the MCC and MNC a to construct the pseudonym. Similarly, the HN would need to compute the hash digest (of $m$ bits) of all the IMSIs (of all the valids subscribers) with the salt using the respective master keys. If the salt is chosen by the UE based on an agreed scheme (e.g., the salt is the current date), then the HN has to compute hash digest for each subscriber according to that scheme (e.g., once in a day) and store in a hash table. However, since $m$ is only around $34$ bits, there would have many collisions --- mutiple IMSIs will be hashed to the same pseudonym. The HN would consult with the hash table when it receives a pseudonym. Due to many collisions in the hash table, pseudonyms would be ambiguous.

\section{Analysis of Our Solution} \label{sec:analysis}

In our solution, the pseudonyms are delivered to the UE with confidentiality protection using the key $\kappa$. Hence, an IMSI catcher cannot know a pseudonym before the UE uses it. This provides unlinkability across the pseudonyms. Once a UE switches to use a new pseudonym, the UE appears as a new (previously unknown) user in the network. Same pseudonym may be transmitted many times in different $\RAND$. However, the challenge still remains fresh and random because of the randomly chosen $l$-bit long $\salt$; the value of $l$ can be as long as $68$ bits. Since a user keeps using the same pseudonym until it obtains a new pseudonym, the UE is exposed to an active IMSI catcher for the time. This exposure mostly coincides with the already existing exposure of the GUTI. In the rest of this section we analyze our solution from important aspects of using IMSI-like pseudonyms in mobile networks: (i) synchronization of pseudonyms, (ii) LI and patching, (iii) billing and charging, and (iv) performance overheads.

\subsection{Synchronization} \label{sec:synchronization_alanlysis} 
\subsubsection{What is Desynchronization?} We say that a UE is desynchronized with the HN if the pseudonyms $\pone$ and $\ptwo$ of the user in the UE side are no more associated with the same user in the HN side. In other words, a UE is desynchronized with the HN if the following condition holds. 

$$(\pone,\done),(\ptwo,\dtwo) \notin \{(\pc,\dc),(\pn,\dn),(\pf,\df)\} \cup \Poldhn$$ 

The UE is not allowed to use any other pseudonyms than $\pone,\ptwo$ in response to an IMSI inquiry from an LTE, 3G or GSM SN. Consequently, when the UE responds to an IMSI inquiry with $\pone$ or $\ptwo$, the HN would fail to retrieve the correct IMSI. As a result, the subsequent AKA would fail and the UE will not be able to join the network.

\subsubsection{Can Desynchronization Happen?} If both the HN and UE function correctly desynchronization can not happen. We will present our argument in this section. In principle, desynchronization may happen if one of the following cases happen:
\begin{enumerate}
    \item If UE accepts a pseudonym that is not generated in the HN \label{cases:case1}
    \item If HN deletes $(\pone,\done),(\ptwo,\dtwo)$ from $\{(\pc,\dc),(\pn,\dn),(\pf,\df)\} \cup \Poldhn$ \label{cases:case2}
    \item If UE accepts a pseudonym that was generated for the user in the HN but the HN has already deleted the pseudonym \label{cases:case3}
\end{enumerate}

First, we discuss Case \ref{cases:case1}. Since a pseudonym is embedded in the $\RAND$, the integrity of the pseudonym is protected if the integrity of the $\RAND$ is protected. The integrity of the $\RAND$ with the help of the $\MAC$ that is a part of the $\AUTN$. So, the UE would never accept a pseudonym that was not generated in the HN; consequently, Case \ref{cases:case1} can never happen.

Second, we discuss Case \ref{cases:case2}. The HN never deletes $(\pc,\dc),(\pn,\dn),(\pf,\df)$. The HN may delete a pseudonym $(\pis,\di)$ from $\Poldhn$ if the HN has decrypted a $\deltamin$ from a valid SUCI such that $\di < \deltamin$; see Step 6 of 5G AKA based solution in Section \ref{sec:5G_AKA_based_solution}. Since, the UE would never deletes $(\pone,\done),(\ptwo,\dtwo)$, the value of $\deltamin$ would be at most $\done$. This implies $\deltamin \leq \done < \dtwo$. Consequently, the HN would not delete $(\pone,\done),(\ptwo,\dtwo)$ from $\Poldhn$. 




Third, we discuss Case \ref{cases:case3}.  As discussed in Case \ref{cases:case2}, if pseudonym $(\pis,\di)$ is deleted from $\Poldhn$, then $\dtwo$ would be greater than $\di$, because $\di < \deltamin \leq \done < \dtwo$. Remember that a UE would accept a pseudonym $(\pis,\di)$ only if $\di > \dtwo$. Thus $(\pis,\di)$ would never be accepted by the UE.

\subsubsection{Resynchronization.} However, if due to some unlikely errors in UE and HN, desynchronization happens, our solution can bring back a desynchronized user into a synchronized state automatically just by connecting to a 5G SN. This is because a 5G UE does not need a valid pseudonym to participate in a 5G AKA, and the UE would get a valid pseudonym if it can participate in a valid 5G AKA. This is a major advantage because, as discussed in \cite{ICISS2017}, such a desynchronized user would otherwise have to go to a mobile network operator's shop to change the UICC. 

Also, if $\dtwo$ in the UE gets corrupted and becomes larger than $\df$, then the UE would not be able to accept a new pseudonym; see Algorithm \ref{alg:conctruct_rand_lte},\ref{alg:extract_pseudonym} and \ref{alg:generate_rand_5G}. If $\dtwo$ becomes a large enough value, then the UE might not accept new pseudonyms anymore. However, the 5G UE is able to resynchronize with HN even if such a corruption happens; just by connecting to a 5G SN using SUCI and running a 5G AKA. 

Since the UE would embed $\deltamax$ (which is equal to $\dtwo$) in the SUCI message, the 5G HN would know (Algorithm \ref{alg:generate_suci}) if such a corruption has happened, see Algorithm \ref{alg:generate_rand_5G}. To help the UE fixing the corruption, the 5G HN would set the ECF in the $\RAND$, see Algorithm \ref{alg:generate_rand_5G}. The UE decrypts the $\RAND$ and extracts the pseudonym $(p,d)$. If the ECF is set, the UE would accept the pseudonym even though $d < \dtwo$, see Algorithm \ref{alg:extract_pseudonym}. Since, acceptance of such a pseudonym may break the consistency with other pseudonyms in the UE, the UE deletes all other pseudonyms it has. Pseudonyms $\pone,\ptwo$ holds the same value as $p$; $\done$ gets the value $d-1$ and $\dtwo$ gets $d$. Thus the UE goes back to a state similar to the beginning.


\subsection{Lawful Interception and Patching} \label{sec:lawful_intercpetion_patching}
Lawful Interception (LI) involves selectively intercepting communications of individual subscribers by legally authorized agencies. The agency typically provides the long-term identifier of the target UE or service to the network operator, which must have the means to intercept communications of the correct target based on long-term or permanent identifiers associated with that target. There is a requirement that a network operator is able to intercept without the need to rely on another network operator or jurisdiction. In particular, an SN does not need to share LI target identities of roaming UE with an HN and vice versa \cite{TS33106,TS33126}.  

In order to support LI in Release 15, where the UE encrypts its identity using the public key of the HN, (1) the HN provides long-term identifier of the UE to the SN during authentication and key agreement procedure; and (2) both UE and SN use long-term identifier of the UE as one of the inputs to the derivation of master session key. The visited network gets the long-term identifier of the UE as a result of (1), while (2) ensures that the UE and visited network can communicate only if both use the same long-term identifier of the UE.   

In order to support LI in our pseudonym-based solution, we propose to add (1), and possibly also (2), into the existing legacy core network elements (MME and HSS in LTE; MSC, SGSN, and HLR in 3G/GSM) by software upgrade. The scope of this software upgrade appears to be much smaller, compared to adapting Release 15 solution into the existing legacy networks. This is mainly because adapting Release 15 solution to legacy networks is likely to impact also the radio access networks: the format of encrypted identifier in Release 15 is quite different from the cleartext IMSI, while in our solution the pseudonym is in the format of IMSI.

The LI feature (1) can be implemented in our solution as follows: in the AV request, the SN informs the HN that it is patched and when the HN returns the AV, it includes the MSIN as part of the AV.

Adding support for (2) is more complicated, especially in the cases of 3G and GSM. In LTE, the SN extracts the MSIN and uses it as an additional input in deriving the master session key and the UE is also informed that the SN is patched by piggybacking on $\RAND$ so that the UE also uses the MSIN in computing the master session key. In 3G and GSM there is no concept of master session key. Therefore, the MSIN has to be used directly in derivation of ciphering and integrity keys. 

Please note that this complication appears also in the case where Release 15 public-key based solution would be adapted for legacy networks. 

Typically, an organization of mobile network operators, like GSMA (GSM Association), could discuss and agree on a deadline in order to provide enough time for operators to patch their networks.

If some legacy network has not completed the patch by the agreed deadline, it is OK for other operators to ignore that failure and start using the pseudonyms-based protection of identity privacy. The LI in the network that is late in patching would not work fully for roaming 5G UEs. 

In the case of a public key-based solution, if some mobile operator is late with a patch, then roaming 5G UEs would not be able to join that network.


\subsection{Charging and Billing}
\label{sec:charging and billing}
Mobile users are charged based on CDRs (call detailed records) that are generated in the serving network. A CDR includes IMSI and records chargeable event (like the service used). The home network then adds charges to the bill of subscriber that is associated with the CDR's IMSI.

After a pseudonym-based solution is adopted, 
a CDR may contain a pseudonym in place of IMSI. (Indeed, if the UE uses pseudonym in IMSI format when it communicates with SN, and the SN does not get the long-term identifier of that UE -- for instance, because it was not patched for LI, then the SN has no other choice but to put the UE's pseudonym into the CDR.)  For this reason the home network has to consult a log of pseudonym allocations when it maps from UE identifiers in received CDRs to the actual subscribers' data \cite{SSR15,ICISS2017}. That log includes the following information: (i) the pseudonym, (ii) the IMSI of the subscriber whom the pseudonym is given to, (iii) the time when HN allocated the pseudonym to the subscriber, (iv) the time when the subscriber started using the pseudonym (i.e., the time of the first successful AKA run of that subscriber with the pseudonym), (v) the time when the UE notified that it is no longer using the pseudonym, and (vi) list of SNs that the user has attached with using this pseudonym. The home network has to maintain the log at least until the billing is settled, and possibly for longer time, to comply with the local authority's guidelines.

In order to keep the billing accurate, a UE that stops using pseudonym $p$ and deletes it, must also stop using the GUTI and the security context associated with $p$. Let us illustrate what may happen otherwise. A UE1 that is visiting SN1, removes a pseudonym $p$ from its $\Poldue$ but keeps the GUTI and the security context. The UE1 also informs the HN (via 5G SUCI) that it has removed the pseudonym $p$. Therefore, the HN sends the pseudonym $p$ to the pool of free pseudonyms; subsequently, $p$ is allocated to another subscriber and delivered to UE2 in SN2. 

If UE1 continues to identify itself in SN1 by the GUTI associated with $p$ as it consumes services, then SN1 will generate a CDR that contains the pseudonym $p$, which is at that time is allocated to UE2. As a result, UE2 will be charged for the service consumed by UE1 in SN1. This may continue until there is a new AKA run between UE1 and SN1 or until the GUTI expires. However, the HN has enough information in the pseudonym allocation log to resolve the correct UE who created the CDR.

\subsection{Performance Overheads}
In our solution the pseudonyms' delivery protocol between UE and HN is piggybacked on existing AKA messages. The structure of AKA messages remains the same, but parts of those messages are constructed and interpreted differently. We list here the additional tasks that have to be done by HN and UE.

Most of those tasks have to be done in the HN: First, the HN has to store a set of in-use pseudonyms per subscriber, maintain a pool of free pseudonyms, and be able to allocate pseudonyms from this pool uniformly at random.

Second, as mentioned in Section \ref{sec:charging and billing}, the HN needs to log  pseudonym assignments for charging purposes. Keeping this log adds to the storage overhead in the HN and computational overhead in the charging system.

Third, the HN has to generate a random $l$-bit  $\salt$ and encrypt the pseudonym and $\salt$ using the symmetric key $\kappa$. The 5G HN has also to verify the MAC of the message that is encrypted into the SUCI. But the overheads due to these symmetric-key cryptographic operations are negligible. 

Fourth, there is one-time, initial provisioning effort  into the USIM of: (i) the initial pseudonyms, and (ii) a policy for forgetting pseudonyms.

On the UE side, one extra decryption is needed to extract the pseudonym embedded in $\RAND$. This symmetric-key decryption has negligible overhead. In addition, UE has to maintain a set of pseudonyms based on the policy provisioned by the HN.

\section{Conclusion} \label{sec:conclusion}
3GPP Release 15, the first release of 5G system, includes protection of long-term identity of mobile users against active IMSI catching. But in a typical case where 5G UE also supports LTE, it is still vulnerable to LTE IMSI catchers. This threat can be mitigated by adopting pseudonyms in the format of IMSI in LTE. 

We propose a solution where the Release~15 mechanism for protecting user identity privacy is used for synchronizing the LTE pseudonyms in the format of IMSI  between the UE and HN, thus making the LTE pseudonyms more robust. 
Our solution can automatically bring a desynchronized user back to synchronized state just by connecting to a 5G network. For LI purpose, the required patching effort in the legacy SNs is reasonable. All in all, pseudonym-based solution is a potential candidate for confounding IMSI catchers in legacy networks. 

Questions for future study include the following: (i) Is it a good idea to encrypt the pseudonyms always with the same key? (ii) What data structure can be used to maintain the pool of free pseudonyms? (iii) What part of the solution has to be implemented in the USIM and what part in the ME? Capping the size of the set $\Poldhn$, when the UE is connecting only with LTE SNs, is also a subject of further study. The resilience of pseudonym-based solution to computational errors in HN, SN, or UE would also be an important area of investigation.


\bibliographystyle{splncs}
\bibliography{ref}{}



\begin{subappendices}
\renewcommand{\thesection}{\Alph{section}}%

\section{Summary of Notation} \label{appendix:notations}

\begin{longtable}{p{1.75cm}p{10cm}}
$\AUTN$ & authentication token used to verify the integrity of the $\RAND$; computed by the HN; sent to the SN as part of the AV; used in both LTE AKA and 5G AKA\\
AV & authentication vector; a bunch of information that the HN sends to delegate the authentication to the SN\\
CDR & call detailed record\\
$d_i$ & the counter value (time stamp) associated with pseudonym $p_i$ by the HN\\
$\deltamin$ & the smallest counter of all the pseudonyms available in the UE. The UE embeds it in the SUCI.\\
$\deltamax$ & it is always set to the current value of $\dtwo$. The UE embeds it in the SUCI.\\
$\ECF$ & error correction flag, piggybacked on the random challenge $\RAND$. the 5G HN sets this flag to inform the UE that the UE's pseudonym state is corrupted\\
5G AKA & an authentication protocol used in 5G\\
GSM & Global System for Mobile Communications\\
HN & Home network; the UE is subscribed with this network\\
IMSI & international mobile subscriber identity\\
$\kappa$ & a symmetric key shared only between the USIM and the HN\\
$\K$ & the symmetric key, also known as master key, shared only by the USIM of a user and the HN\\
$\KASME$ & anchoring key -- subsequent encryption and integrity protection keys are derive from it; computed by the HN and sent to the SN as part of the AV\\
$l$ & bit length of the $\salt$ that is used as a part of the plaintext that is encrypted into $\RAND$\\
LTE AKA & an authentication protocol used in LTE network\\
LU & location update; when a user attaches to an LTE SN for the first time, after the authentication succeeds, the SN send an LU message to the HN \\
$m$ & number of bits used to generate the randomized decimal digits to construct pseudonyms; $m \leq 40$\\
ME & mobile equipment, usually a mobile phone\\
$\pc,\pn,\pf$ & the pseudonyms in the HN; $\pc$ for the current pseudonym, $\pn$ for new pseudonym and $\pf$ for the future pseudonym; ideally expectation is $\pone = \pc, \ptwo = \pn$ or $\pone = \pn, \ptwo = \pf$\\
$\Poldhn$ & the set of pairs $(\pis,\di)$ maintained by the HN for a specific UE\\
$\pone,\ptwo$ & the pseudonyms the UE is currently using \\
$\Poldue$ & the set of pairs $(\pis,\di)$ maintained by the UE\\
$q$ & the pseudonym that a UE uses to identify itself \\
$\RAND$ & is the 128-bit long random challenge; chosen/constructed by HN; sent to SN as part of the AV; SN forwards it to the UE; used in both LTE AKA and 5G AKA\\
$\RESLTE$ & response sent by the UE to the random challenge $\RAND$ in LTE AKA\\
SN & serving network; the UE connects with this network\\
$\SQN$ & the sequence number used by the USIM and the HN, both in LTE AKA and 5G AKA to defeat replay attack.\\
UE & user equipment (USIM + ME)\\
UICC & universal integrated circuit card\\
USIM & universal subscriber identity module\\
$\XRES$ & expected response (to $\RAND$) from the UE; computed by the HN; sent to the SN as part of the AV; used in LTE AKA\\
$\XSQN$ & the sequence number sent by the HN to the UE\\
\end{longtable}

\section{Algorithms} \label{appendix:algorithms}

\begin{algorithm}
\caption{Construct $\RAND$ for LTE AKA in HN}
\label{alg:conctruct_rand_lte}
\begin{algorithmic}[1]
\Require $q, \left(\text{IMSI}, \kappa, (\pc,\dc),(\pn,\dn), (\pf,\df)\right),\COUNTER$; where $q$ is the pseudonym received in the AV request, the vector following $q$ contains the relevant records of the user in the subscription database s.t., $q \in \{\text{IMSI},\pc,\pn,\pf\} \textbf{ or } (q,*) \in \Poldhn $, and $\COUNTER$ is a non decreasing counter that the HN maintains. $\COUNTER$ increases once in a configured time interval.
\Ensure $\RAND$
        \If{$p_f = \Null$ } \label{state:if_pf_is_null}  
            \State allocate $\pf \in \{0,1\}^{m}$ randomly from the pool of free pseudonyms
            \State $\df \leftarrow \COUNTER$
        \EndIf
        \State choose $\salt \in \{0,1\}^l$ randomly
        \State $(p,d) \gets (\pf,\df)$
        \State $\ECF \gets 0$ \Comment{this flag might be set to $1$ by a 5G HN to indicate an error}
        \State $\RAND \gets E_{\kappa}\left(p,d,\ECF,\salt \right)$ \label{state:creating_rand_by_encryption}
    \State \textbf{return} $\RAND$
\end{algorithmic}
\end{algorithm}

\begin{algorithm}
\caption{Update Pseudonyms in UE}
\label{alg:extract_pseudonym}
\begin{algorithmic}[1]
\Require $\RAND$, $\pone,\ptwo$
\Ensure updated pseudonym states in the UE
    \State extract $p,d,\ECF, \salt$ by decrypting $\RAND$ using key $\kappa$ \label{state:decrypting_rand}
        \If{$\ECF = 1$} \label{state:ue_knows_error_happened} \Comment{it means, the UE has unreasonably large $\dtwo$}
            \State empty the set $\Poldue$ 
            \State $(\pone,\done),(\ptwo,\dtwo) \gets  (p,d-1), (p,d)$
            \State \textbf{return} \label{state:ue_knows_error_happened_ended}
        \EndIf
        \If{$d > \dtwo$} \label{state:p_not_in_p1_p2}
            \State $\Poldue \leftarrow \Poldue \cup \{(\pone,\done)\}$ \label{state:upue_inserting_into_poldue}
            \State $(\pone,\done),(\ptwo,\dtwo) \gets  (\ptwo,\dtwo), (p,d)$ \label{state:upue_shifting_pseudonyms}
        \EndIf \label{state:end_inner_if}
\end{algorithmic}
\end{algorithm}

\begin{algorithm}
\caption{Update Pseudonyms in HN after LTE AKA}
\label{alg:update_pseudonym_hn}
\begin{algorithmic}[1]
\Require $q, \left(\Poldhn, (\pc,\dc),(\pn,\dn),(\pf,\df)\right)$; where $q$ is the pseudonym received in the AV request, the vector following $q$ contains the records of the user in the subscription database s.t., $q \in \{\pn,\pf\}$
\Ensure updated pseudonym states in the HN
        \If{$\pf \neq \Null $}
            \State $\Poldhn \leftarrow \Poldhn \cup \{(\pc,\dc)\}$ \label{state:uphn_inserting_into_poldhn}
            \State $(\pc,\dc),(\pn,\dn),(\pf,\df) \gets (\pn,\dn),(\pf,\df),(\Null, \Null)$ \label{state:uphn_shifting_pseudonyms}
        \EndIf
\end{algorithmic}
\end{algorithm}

\begin{algorithm}
\caption{Generate SUCI}
\label{alg:generate_suci}
\begin{algorithmic}[1]
\Require $\MSIN, \K, \delta_{\text{min}}, \delta_{\text{max}}, \HNID, \pk, \HNPKI, \SPSI$; where $\delta_{\text{min}}$ is the counter of the earliest pseudonym in $\Poldue$; $\delta_{\text{max}}$ is $\dtwo$;  $\HNID$ is the HN ID, usually $\MCC||\MNC$; $\pk$ is the public key of the HN; $\HNPKI$ is the public key identifier of the HN; $\SPSI$ is the SUPI protection scheme identifier
\Ensure SUCI
    \State $T \gets \MAC_{\K}\left(\MSIN||\delta_{\text{min}}||\delta_{\text{max}}\right)$ \label{algo_state:suci_generation_T}
    \State $\SUCI = \HNID||\HNPKI||\SPSI||\Encpk_{\pk}\left(\MSIN||\delta_{\text{min}}||\delta_{\text{max}}||T\right)$ \label{algo_state:suci_generation_encryption}
        \State \textbf{return} $\SUCI$
\end{algorithmic}
\end{algorithm}

\begin{algorithm}
\caption{Construct $\RAND$ for 5G AKA}
\label{alg:generate_rand_5G}
\begin{algorithmic}[1]
\Require $\sk,\SUCI,\kappa,(\pf,\df),\COUNTER$; where $\sk$ is the private key of the HN, $\SUCI$ is received from the SN in the AV request, $\kappa$ is the subscriber-specific key to encrypt pseudonyms, $\left( \pf,\df \right)$ is the subscriber's pseudonym and its counter that would be embedded in $\RAND$
\Ensure $\RAND$
    \State extract $\text{MSIN},\delta_{\text{min}}, \delta_{\text{max}}, \text{ and } T$ by decrypting $\SUCI$ using secret key $\sk$
    \label{algo_state:suci_extraction}
            \If{$T = \MAC_{\K}\left( \MSIN||\delta_{\text{min}}||\delta_{\text{max}}\right)$} \label{algo_state_T_verification}
                \If{$\pf = \Null$} \label{algo_state_constructing_rand_5G_AKA_pf_null_check}
                    \State allocate $\pf \in \{0,1\}^{m}$ randomly from the pool of free pseudonyms \label{algo_state:allocating_pf_5G}
                    \State $\df \leftarrow \COUNTER$ \label{algo_state:setting_df_5G}
                \EndIf
                \State $\ECF \gets 0$ \label{algo_state:setting_ECF_start}
                \If{$\deltamax > \df$} \Comment{It means the UE has unreasonably large $\dtwo$}
                    \State $\ECF \gets 1$ 
                \EndIf \label{algo_state:setting_ECF_end}
                \State choose $\salt \in \{0,1\}^l$ randomly \label{algo_state:choosing_salt_5G}
                \State $(p,d) \gets (\pf,\df)$ \label{algo_state:setting_p_d_5G}
                \State $\RAND \leftarrow E_{\kappa}\left(p,d,\ECF,\salt\right)$ \label{algo_state:encrypting_into_rand_5G}
            \State \textbf{return} $\RAND$
        \EndIf

\end{algorithmic}
\end{algorithm}

\begin{algorithm}
\caption{Update Pseudonyms in HN after 5G AKA}
\label{alg:update_pseudonym_hn_5G}
\begin{algorithmic}[1]
\Require $p, \left(\Poldhn, (\pc,\dc),(\pn,\dn),(\pf,\df)\right)$; where $p$ is the pseudonym that was embedded in the $\RAND$ of the 5G AKA in question, the vector following $p$ contains the relevant records of the user participated in the AKA.\\
\textbf{Output:} updated pseudonyms in the HN
        \If{$\pf = p$}
            \State $\Poldhn \gets \Poldhn \cup \{(\pc,\dc)\}$
            \State $(\pc, \dc),  (\pn,\dn), (\pf, \df) \gets (\pn,\dn), (\pf, \df), (\Null, \Null)$
        \EndIf
\end{algorithmic}
\end{algorithm}

\end{subappendices}


\end{document}